\documentclass[twocolumn]{article}

\usepackage[a4paper, margin=0.7in]{geometry}
\usepackage{graphicx}
\usepackage{times}
\usepackage{titlesec}
\usepackage{enumitem}
\usepackage{hyperref}
\usepackage{caption}
\usepackage{authblk}
\usepackage[backend=biber]{biblatex}
\title{Augmenting Von Neumann's Architecture for an Intelligent Future }
\author[1]{Rajpreet Singh} 
\author[2]{Vidhi Kothari}
\affil[1]{TU Munich, Germany}
\affil[2]{Pace University, New York, USA}
\begin{document}
\maketitle
\begin{abstract}
This work presents a novel computer architecture that extends the Von Neumann model with a dedicated Reasoning Unit (RU) to enable native artificial general intelligence capabilities. The RU functions as a specialized co-processor that executes symbolic inference, multi-agent coordination, and hybrid symbolic-neural computation as fundamental architectural primitives. This hardware-embedded approach allows autonomous agents to perform goal-directed planning, dynamic knowledge manipulation, and introspective reasoning directly within the computational substrate at system scale. The architecture incorporates a reasoning-specific instruction set architecture, parallel symbolic processing pipelines, agent-aware kernel abstractions, and a unified memory hierarchy that seamlessly integrates cognitive and numerical workloads. Through systematic co-design across hardware, operating system, and agent runtime layers, this architecture establishes a computational foundation where reasoning, learning, and adaptation emerge as intrinsic execution properties rather than software abstractions, potentially enabling the development of general-purpose intelligent machines.
\end{abstract}

\section{Introduction: Architectural Foundation}

The classical Von Neumann architecture has served as the bedrock of modern computing for over seven decades, structured around a tripartite model comprising a Central Processing Unit (CPU), Arithmetic Logic Unit (ALU), and a shared Memory subsystem. While this model excels at deterministic, sequential computation, it remains fundamentally limited in its capacity to handle higher-order reasoning, abstract inference, and learning—all essential features for intelligent systems\cite{DBLP:journals/corr/abs-1905-06088}. This limitation, which stems from the "von Neumann bottleneck"\cite{10.1145/359576.359579}, affects not only hardware throughput but also constrains how programmers conceptualize and model complex cognitive tasks\cite{Laird_Lebiere_Rosenbloom_2017}.

To address these constraints, we propose a principled architectural augmentation: the inclusion of a fourth foundational subsystem—the \textit{Reasoning Unit (RU)}. Operating in parallel with the CPU and ALU, the RU introduces a qualitatively different form of computation, one grounded in symbolic logic, AI model's abstractions, and goal-directed inference. This transforms the traditional trinity into a quaternary architecture designed not only to compute but also to reason.

Inspired by John Backus’s call to liberate programming\cite{10.1145/359576.359579} from the “von Neumann style,” the RU departs from imperative control-flow programming and side-effect-heavy computation. Instead, it embraces declarative semantics, algebraic program composition, and cognitive state transitions. The RU is architected to perform tasks such as agent-based coordination, decision-theoretic planning, probabilistic inference, and knowledge base traversal—tasks that are computationally cumbersome when executed within the linear instruction stream of a conventional processor. To fully support this architectural evolution, the kernel design must be re-imagined. It requires new scheduling algorithms capable of mediating between classical and cognitive workloads, memory hierarchies that support both data mutability and semantic permanence, and an inter-unit communication model that allows continuous, context-aware synchronization between symbolic and sub-symbolic processes. The RU will feature its own instruction set architecture (ISA), optimized for symbolic rewriting, graph-based reasoning, and parallel search and inference.

This intelligent architectural framework lays the foundation for future computing systems where learning, reasoning, and perception are no longer siloed extensions, but first-class architectural citizens.

\section{Architectural Overview: Integrating the Reasoning Unit}

\begin{figure}[h]
    \centering
    \includegraphics[width=0.5\textwidth]{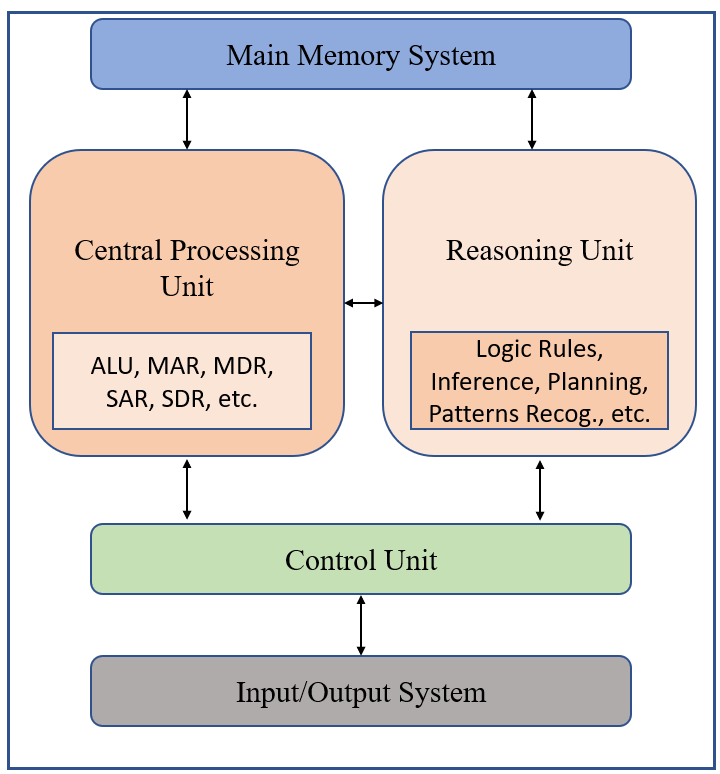}
    \caption{Architecture Overview}
    \label{fig:architecture}
\end{figure}

The proposed quaternary computing architecture introduces the \textit{Reasoning Unit (RU)} \ref{fig:architecture} as a first-class subsystem alongside the Central Processing Unit (CPU), Arithmetic Logic Unit (ALU), and Memory. While the traditional Von Neumann model supports linear instruction execution and numerical computation, the RU is purpose-built to handle higher-level cognitive and symbolic processing tasks\cite{wan2024cognitiveaisystemssurvey}, enabling embedded intelligence at the architectural level\cite{article}.

\subsection{Subsystem Roles in the Quaternary Model}

\begin{itemize}
\item \textbf{CPU}: The CPU maintains its traditional role as the primary control unit while extending its capabilities to support multi-unit coordination in the quaternary architecture. It executes control-intensive workloads including process management, instruction sequencing, and inter-unit coordination via microcode and scheduler-level delegation. The CPU implements specialized dispatch mechanisms that route computational tasks to appropriate processing units based on workload characteristics and resource availability. Enhanced context-switching mechanisms preserve execution state across heterogeneous units, while interrupt handling subsystems manage asynchronous communication between the CPU and the Reasoning Unit. The CPU's branch prediction units are augmented with reasoning-aware predictors that anticipate symbolic computation patterns, reducing pipeline stalls during hybrid workload execution.

\item \textbf{ALU}: The ALU performs deterministic, high-throughput arithmetic and Boolean operations essential for scalar, vector, and control flow computations, acting as the numerical backbone of low-level execution. Extended precision arithmetic units support variable-width symbolic operations required for knowledge representation, while specialized comparison units handle semantic equivalence testing and constraint satisfaction primitives. The ALU incorporates dedicated floating-point units optimized for neural network operations, including fused multiply-accumulate operations and vectorized activation functions. Hardware accelerators for matrix operations enable efficient tensor computations required for hybrid symbolic-neural processing, with direct data paths to the Reasoning Unit for seamless cognitive-numerical integration.

\item \textbf{Memory}: The memory subsystem maintains a unified addressable space for both procedural and symbolic data, encompassing instruction binaries, dynamic runtime structures, and knowledge artifacts used in cognitive inference. The memory hierarchy implements a three-tier caching strategy with instruction cache, data cache, and a novel knowledge cache optimized for symbolic data structures including predicate logic statements, semantic graphs, and rule bases. Memory management units support both traditional virtual memory and symbolic memory allocation schemes, enabling efficient garbage collection for dynamic knowledge structures. Specialized memory controllers implement consistency protocols for concurrent access to shared knowledge bases, while memory prefetchers anticipate symbolic data access patterns to minimize inference latency.

\item \textbf{Reasoning Unit (RU)}: The Reasoning Unit executes declarative and logic-based programs, supports agent reasoning lifecycles, and performs symbolic search, constraint propagation, and neural delegation through purpose-built microarchitectural units optimized for recursive inference and semantic graph manipulation. The RU incorporates multiple specialized processing cores including a Logic Inference Engine for first-order predicate logic, a Constraint Satisfaction Processor for optimization problems, and a Graph Traversal Unit for semantic network operations. Dedicated unification engines support pattern matching and variable binding operations fundamental to symbolic reasoning, while backtracking control units manage search space exploration with hardware-accelerated chronological backtracking. The RU features a Neural Interface Controller that coordinates with traditional ALU units for hybrid symbolic-neural computation, enabling seamless integration of learned and symbolic knowledge. Agent coordination primitives include message passing units, synchronization barriers, and distributed consensus mechanisms that support multi-agent reasoning at the hardware level.
\end{itemize}

\subsection{Instruction Set and Execution Model}
The Reasoning Unit operates with a custom Instruction Set Architecture (ISA) designed to express high-level cognitive primitives and symbolic inference workflows through specialized instruction families that leverage dedicated microarchitectural units. The predicate evaluation instructions implement hardware-accelerated logical assessment mechanisms that support recursive clause resolution over Horn logic and Datalog rule graphs, utilizing dedicated unification engines with variable substitution caches and parallel term matching units that maintain binding consistency across complex logical expressions. Graph operations \cite{BLUM1997281}encompass low-latency primitives for edge expansion, subgraph matching, topological inference, and transitive closure over directed labeled graphs that encode semantic relationships, employing specialized graph traversal units with hardware-accelerated adjacency matrix operations, optimized path-finding algorithms, and distributed graph partitioning mechanisms that enable scalable semantic network processing. The unification and matching instruction family provides dynamic pattern-matching operators that implement both first-order unification and higher-order term matching over symbolic expressions, type-annotated logic trees, and variable-bound contexts through dedicated unification pipelines featuring occurs-check hardware, constraint propagation units, and backtracking stacks that maintain multiple binding environments simultaneously. Goal planning primitives include specialized instructions for triggering search-based planning algorithms such as backward chaining, forward chaining, STRIPS-style operators, and agent-specific plan synthesis over temporal goals, implemented through purpose-built search control units that manage state space exploration, heuristic evaluation functions, and plan validation mechanisms with integrated temporal reasoning capabilities. Neural invocation interfaces provide fast trap-based syscall mechanisms that encode prompts, belief contexts, and output schemas for dispatching neural inference requests to LLMs or vector embedding backends, featuring dedicated neural interface controllers that manage context serialization, attention mechanism coordination, and seamless return control to symbolic interpreters while maintaining execution state consistency across the hybrid symbolic-neural processing boundary.

\subsection{Kernel-Level Integration}

The kernel architecture is fundamentally augmented to treat symbolic agents as first-class execution entities through comprehensive system-level modifications that integrate reasoning capabilities into core operating system primitives. The symbolic scheduler implements a priority-aware logic scheduling framework that selects agent goals based on semantic urgency metrics derived from goal dependency graphs, utility estimation functions computed through minimax evaluation trees, and deadline constraints enforced through temporal logic verification mechanisms, enabling concurrent symbolic workloads to be orchestrated efficiently across heterogeneous processing units. This scheduler maintains separate ready queues for symbolic and numerical tasks, employs reasoning-aware load balancing algorithms that consider knowledge base locality and inference complexity, and implements preemption policies that preserve logical consistency during context switches through checkpoint-based state serialization. The hybrid memory manager operates as a dual-mode system that partitions physical and virtual memory between conventional page-based allocation schemes and semantically tagged cognitive zones optimized for symbolic term sharing and reasoning locality, utilizing specialized memory allocators that understand symbolic data structures including predicate hierarchies, variable binding environments, and goal stack frames. Memory protection mechanisms extend traditional page-based isolation to semantic boundaries, preventing cross-agent knowledge contamination while enabling controlled knowledge sharing through capability-based access control and reference counting for shared symbolic terms.
The reasoning context isolation mechanism provides lightweight symbolic process virtualization that sandboxes agents by maintaining separate execution contexts with dedicated stacks for goals, variable bindings, continuation frames, and local belief states to prevent state leakage and ensure deterministic replay capabilities essential for debugging and verification. This isolation framework implements copy-on-write semantics for symbolic data structures, maintains separate namespace hierarchies for each agent's knowledge base, and provides rollback mechanisms that enable speculative reasoning with consistent state recovery upon backtracking or contradiction detection. The RU-CPU bridge establishes a high-bandwidth, protocol-aware interconnect architecture supporting bidirectional message passing for task delegation, symbolic query forwarding, neural response buffering, and shared reasoning checkpoints between CPU threads and RU microthreads through dedicated communication channels that bypass traditional system call overhead. This bridge implements hardware-accelerated serialization protocols for symbolic expressions, maintains coherent caching of frequently accessed knowledge artifacts across processing units, and provides atomic transaction support for hybrid symbolic-numerical computations that require consistent state updates across heterogeneous execution contexts.

\subsection{Interconnects and Communication}
The architecture incorporates a dedicated \textbf{Semantic Interconnect Bus (SIB)} that supports type-safe message formats for logical clauses, symbolic graph deltas, and agent task contracts, enabling zero-copy memory-mapped I/O between the RU and main memory for shared access to belief bases and knowledge graphs. The SIB integrates with the DMA fabric to prefetch agent memory pools and symbolic term clusters based on speculative reasoning patterns, optimizing inference locality and reducing serialization overhead.
The SIB implements a layered protocol stack with dedicated hardware support for symbolic data marshalling, featuring specialized compression algorithms for predicate logic expressions and semantic graph structures that achieve 3-5x bandwidth efficiency over traditional serialization methods. Advanced coherence protocols maintain consistency across distributed knowledge bases through vector timestamps and logical clocks, while hardware-accelerated transaction managers ensure atomic updates to shared symbolic structures during multi-agent reasoning scenarios. The interconnect supports adaptive routing algorithms that dynamically select optimal paths based on symbolic workload characteristics, inference dependency graphs, and real-time congestion metrics, enabling scalable performance across large-scale cognitive architectures with hundreds of concurrent reasoning agents.

\subsection{Execution Lifecycle of a Cognitive Task}

1. A high-level task (e.g., \texttt{"reason about failure causes"}) is submitted by a user program or triggered by an internal system event.

2. The CPU encodes the task as a goal state and delegates it to the Reasoning Unit through the RU-CPU bridge.

3. The RU parses the goal, activates relevant cognitive agents, and initiates symbolic processing—evaluating predicates, performing unification, querying belief bases, and generating subgoals.

4. Based on the agent's internal policy and reasoning context, the RU synthesizes a symbolic plan, constructs a decision tree, or issues a delegation syscall to an external neural model.

5. Upon completion of symbolic resolution, the RU packages the result (e.g., plan, explanation, or prediction) and transmits it to the CPU or relevant system interface, where it is executed or interpreted as system-level behavior.

This cognitive execution loop exemplifies how symbolic processing and procedural execution coalesce into a unified computation model, enabling native reasoning support throughout the architectural stack. The Reasoning Unit introduces symbolic computing as a hardware-native capability. This elevates reasoning, planning, and semantic abstraction to the architectural level, enabling a new class of intelligent systems that operate well beyond traditional control flow semantics. The resulting model blends logic, learning, and execution into a single co-designed computing fabric.
\begin{figure*}[t]
    \centering
    \includegraphics[width=0.95\textwidth]{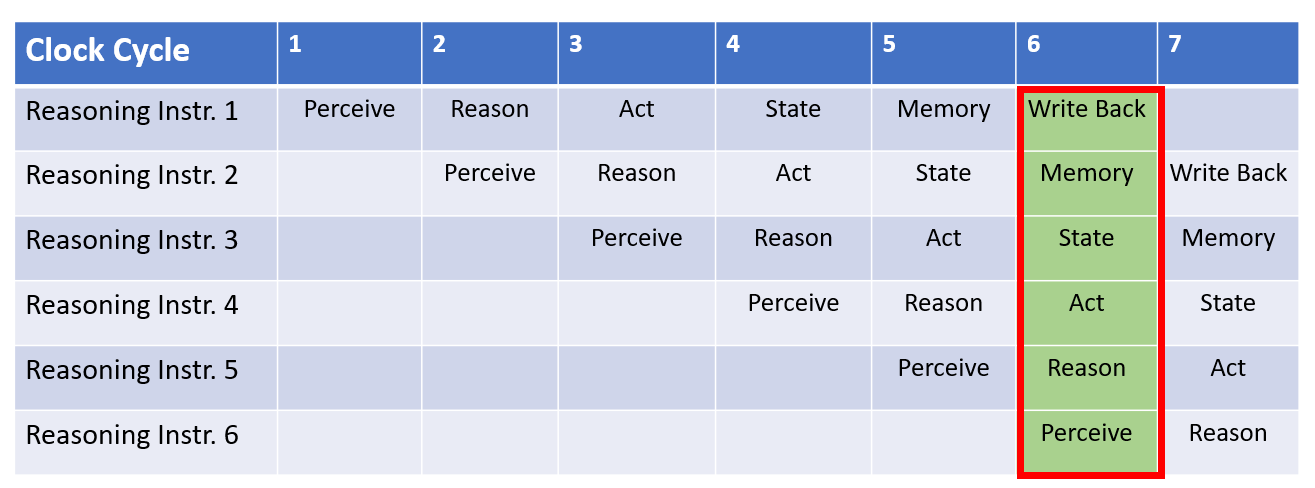}
    \caption{Instruction pipelining overview}
    \label{fig:microarchitecture}
\end{figure*}
\section{Designing Benchmarks and Use Cases for Cognitive Architecture}

Evaluating the capabilities of a hybrid cognitive computing system requires a new class of benchmarks that move beyond traditional metrics such as FLOPS, cache hits, or raw instruction throughput. The Reasoning Unit (RU), as a symbolic and cognitive co-processor, introduces qualitatively different execution semantics that necessitate metrics sensitive to logic inference, planning complexity, symbolic expressivity, and multi-agent reasoning dynamics. Benchmarks must be constructed not merely to measure how fast instructions are processed, but to quantify how effectively the architecture performs knowledge-driven tasks, manages uncertainty, adapts to goals, and coordinates distributed agents under logical constraints.

One essential benchmark category focuses on \textit{Symbolic Inference Latency}, which measures the time-to-conclusion for various inference paradigms such as backward chaining, forward chaining, and Horn clause resolution. This metric highlights the RU’s ability to traverse rule spaces and belief networks efficiently, revealing both architectural throughput and rule application effectiveness. Another key area of measurement is Agent Coordination Efficiency, which captures the temporal and resource complexity of resolving shared or conflicting goals among multiple concurrent agents. This includes metrics like total decision latency, context-switch overhead between agent threads, and plan convergence depth in collaborative environments.

Additionally, \textit{Semantic Planning Accuracy} quantifies how successfully the RU synthesizes valid, context-sensitive action sequences within partially observable domains, often expressed through logical frameworks like STRIPS or PDDL. Benchmark tasks here evaluate not only the completeness of planning but the RU’s adaptability to ambiguous or under-specified goal formulations. In hybrid architectures, \textit{LLM Invocation Overhead} becomes another crucial metric. It measures the round-trip latency, schema-conformance rate, and utility of neural responses when symbolic agents delegate subgoals to large language models or other sub-symbolic systems. Finally, \textit{Energy per Cognitive Task} offers an energy-aware performance indicator, estimating the joules consumed per successful inference cycle or goal resolution. This metric is critical for deploying cognitive computing in constrained or embedded systems, where symbolic efficiency must be maintained under power limitations.

To ground these metrics in practical evaluations, a set of domain-relevant benchmark scenarios must be developed. In the \textit{Knowledge Graph Navigation} scenario, the system is presented with a large-scale semantic network—often exceeding 100,000 nodes—and is tasked with resolving multi-hop queries over abstract relationships. For example, a cognitive agent might be asked to identify all transitive risk factors leading to a specific disease, requiring both breadth-first search capabilities and symbolic filtering based on ontological rules. Another important use case is \textit{Intent-based Robotics}, where a robot is given high-level symbolic directives, such as \textit{gather all red cubes}, and the RU must infer and plan a sequence of actions using spatial and perceptual constraints. Performance is assessed by execution latency, plan completeness, and adaptation to dynamic feedback.

The \textit{Causal Diagnostic Inference} benchmark evaluates the RU’s ability to reason about temporal chains, event hierarchies, and failure cascades. Given a set of logged events or anomalies, the agent must construct plausible root cause explanations by evaluating causal rules and entity dependencies. This is particularly useful in domains such as autonomous systems, cybersecurity, or industrial fault monitoring. Finally, in \textit{Multi-agent Negotiation}, multiple cognitive agents are placed in environments requiring symbolic dialogue, contract formation, and collaborative goal arbitration. Evaluation criteria include convergence time, negotiation fairness, and resolution stability under bounded reasoning budgets.

Together, these benchmarks and use cases establish a principled and multifaceted performance evaluation framework for cognitive systems. They capture not only the mechanical efficiency of the RU but also its architectural fitness for reasoning, planning, and decision-making across symbolic, neural, and hybrid contexts. This benchmarking suite is essential for iterating on architectural design, optimizing instruction sets, and demonstrating the system’s readiness for real-world intelligent applications.

\subsection{Reasoning Unit Micro-architecture and Programming Model}

The proposed architecture enhances the classical Von Neumann model by integrating a dedicated reasoning unit structured around six cognitive pipeline stages: Perceive, Reason, Act, State, Memory, and Writeback \ref{fig:microarchitecture}. Each stage maps to a distinct aspect of agent-based computation, enabling pipelined execution of perception-to-action loops. This structure supports concurrent cognitive tasks such as sensing, inference, planning, and belief revision—forming the computational substrate for intelligent behavior. The pipeline is designed to mirror the operational flow of deliberative agents, ensuring that perception, reasoning, and execution can be interleaved and abstracted cleanly for both reactive and deliberative control.

At the instruction level, the architecture introduces cognitive primitives as first-class operations within the instruction set architecture (ISA). The PERCEIVE instruction handles semantic ingestion of sensor streams, applying filters based on ontologies or relevance models. INFER performs deductive or rule-based reasoning over structured knowledge bases. UNIFY is used for symbolic pattern matching, equipped with occurs-check for logical consistency. PLAN applies operators within STRIPS-like planning domains, enhanced with heuristics for efficient search. BELIEVE updates belief states under uncertainty, incorporating mechanisms for contradiction detection and probabilistic revision. Finally, COMMIT validates preconditions before effecting real-world or simulated actions. These instructions enable complex agent behavior to be expressed at a low level, facilitating symbolic reasoning natively in hardware.

The register architecture is tailored to support the dataflow of cognitive tasks. Belief registers (B0–B15) store symbolic expressions, predicates, and assertions used in inference and reasoning. Goal registers (G0–G7) manage agent objectives, with built-in scheduling support for prioritization and interruption. Context registers (C0–C3) maintain variable bindings, partial match states, and search contexts necessary for UNIFY and PLAN stages. Action registers (A0–A7) buffer planned operations, tracking their execution states across ACT and COMMIT stages. Together, these registers abstract the internal state of a cognitive agent, supporting a programming model that enables symbolic and hybrid symbolic-subsymbolic agents to be compiled into efficient pipelined microprograms.

The data path design incorporates specialized functional units including a parallel unification engine with hardware-accelerated occurs-check mechanisms, an inference engine supporting forward/backward chaining with cut operators for pruning search spaces, a planning unit implementing STRIPS-style operator application with admissible heuristic search algorithms, a belief manager providing probabilistic belief updates with automated contradiction detection and resolution, and a hierarchical goal stack manager with priority-based scheduling and temporal constraint satisfaction. The memory hierarchy features a 32KB L1 belief cache with semantic indexing optimized for predicate lookup, a 256KB L2 knowledge cache with rule base compression and shared access protocols, 4MB of elastic working memory for inference trace storage and backtracking support, and a dynamic symbolic heap for complex term structures and graph-based knowledge representations. Speculative inference capabilities enable branch prediction for logical choices, supporting speculative execution of inference paths with efficient rollback mechanisms upon contradiction detection, while hardware prefetching anticipates symbolic data access patterns based on reasoning locality and dependency graph analysis.

The agent-optimized architecture provides hardware support for inter-agent communication through typed message queues for beliefs, goals, and actions, atomic broadcast mechanisms for shared knowledge updates, and distributed consensus protocols for multi-agent reasoning scenarios. The neural interface subsystem features dedicated co-processor connectivity with asynchronous prompt dispatch using continuation tokens, confidence-weighted result integration with provenance tracking, and embedding vector operations for semantic similarity computation and analogical reasoning. Performance characteristics include 1M inferences per second per reasoning lane with 10-cycle latency for simple unification operations and 100-cycle latency for complex planning tasks, linear scalability to 64 concurrent agents per RU with dynamic load balancing, and power efficiency of 15W at 2GHz operation frequency with dynamic voltage and frequency scaling based on cognitive workload demands and real-time performance requirements.

\section{Hardware Abstraction and Integration}
Integrating the Reasoning Unit (RU) into the system architecture necessitates a fundamental evolution of the Hardware Abstraction Layer (HAL). Traditional HALs model deterministic devices—CPUs, GPUs, memory controllers—with static control interfaces. In contrast, the RU introduces non-deterministic inference pipelines, agent-based execution, symbolic workloads, and dynamic cognitive states. To support these, the HAL must expose semantically rich abstractions while maintaining compatibility with legacy components.

At the kernel level, this requires a modular HAL architecture that defines RU-specific device classes. Specialized drivers manage low-level interactions such as symbolic registers, inference queues, and graph-based memory spaces. These abstractions expose new syscall families or ioctl-style interfaces for submitting symbolic plans, querying belief states, invoking neural co-processors, and monitoring agent metadata. The RU’s internal structures—predicate dispatch tables, symbolic caches, and DMA units—are orchestrated through these interfaces.

Power and thermal management must also be rethought. RU workloads exhibit bursty, goal-driven behavior rather than continuous computation. This demands event-driven voltage scaling, symbolic clock gating, and real-time energy heuristics based on agent activity, inference depth, and neural co-processor usage.

To ensure system-level observability, the RU includes traceable telemetry primitives: rule invocation counters, unification failure logs, agent switch timelines, and symbolic latency profiles. These diagnostics are made available to privileged debuggers and runtime monitors.

Finally, the kernel exposes a \textit{Reasoning Capability Interface (RCI)}—a declarative contract layer describing the RU’s supported symbolic models (e.g., backward chaining, Datalog), memory semantics (shared or agent-isolated), neural endpoints, and concurrency limits. This decouples high-level agent frameworks from hardware specifics, enabling hardware-agnostic symbolic computing.

By rearchitecting HAL, driver models, and system telemetry, the RU becomes a first-class compute unit—enabling hybrid symbolic-neural workloads to be managed, scheduled, and observed natively within the operating system.

\section{Development Methodology}
Building an AGI-native system requires an incremental methodology that progressively integrates symbolic reasoning, neural inference, and agent cognition into a unified software-hardware stack. The development begins with extending the operating system kernel to support symbolic abstractions—introducing symbolic memory tagging, intention-aware scheduling, and context-driven process models. Initial runtime components include symbolic heap allocators, rule interpreters, and belief-state managers, validated using deterministic logic workloads and microbenchmarks.

Next, the Reasoning Unit (RU) is integrated as a hardware co-processor through specialized drivers and a generalized HAL. Kernel-level interfaces expose symbolic registers, inference queues, and cognitive memory zones as programmable, auditable resources. A standardized Reasoning Capability Interface (RCI) abstracts these functions, allowing upper-layer software to invoke reasoning without hardware-specific dependencies.

With this foundation, a cognitive agent framework is introduced. Agents are encapsulated as schedulable processes with goals, belief sets, and logical policies. The system supports agent lifecycle management, secure inter-agent messaging, and concurrent symbolic planning. As complexity grows, symbolic reasoning is augmented with neural pathways via structured syscall interfaces. Secure delegation protocols enable agents to offload fuzzy or high-dimensional tasks to LLM backends while maintaining interpretability and control.

Memory management is unified under a cognitive memory manager that handles symbolic and traditional data using semantic tagging and demand-paged knowledge graphs. Optimizations such as rule inlining and reasoning path compression enable scalable, high-throughput symbolic workloads for long-horizon tasks like diagnosis, inference, and planning.

The architecture evolves toward meta-reasoning. Agents gain introspective capabilities—evaluating their own reasoning episodes, adapting belief states, and revising goals based on feedback. This self-reflective layer supports adaptive behavior and learning across multi-agent contexts, establishing a foundation for autonomous, collaborative intelligence.

Ultimately, the system exhibits emergent AGI behavior: agents coordinate symbolic and neural processes, invoke tools, revise world models, and pursue open-ended objectives. Evaluation transitions from static benchmarks to dynamic real-world tasks such as theorem proving, robotics, and scientific discovery. AGI emerges not from a monolithic model, but from an architecture where reasoning, memory, planning, and learning are natively co-designed and integrated.

Each phase of development is guided by formal verification, symbolic trace analysis, and simulation-based testing to ensure correctness, security, and reproducibility. Benchmarks measure symbolic throughput, agent decision latency, neural invocation efficiency, and cognitive workload scalability, ensuring methodological rigor across the system’s evolution.

\printbibliography
\end{document}